# Pressure dependence of $T_c$ in the MgB$_2$ superconductor as probed by resistivity measurements

E Saito[1], T Taknenobu[1], T Ito[1], Y Iwasa[1] and K. Prassides[1,2,3]

[1] *Japan Advanced Institute of Science and Technology, Tatsunokuchi, Ishikawa 923-1292, Japan*
[2] *School of Chemistry, Physics, and Environmental Science, University of Sussex, Brighton BN1 9QJ, UK*
[3] *Institute of Materials Science, NCSR "Demokritos", 15310 Ag. Paraskevi, Athens, Greece*

Abstract

High-pressure resistivity experiments were performed on the recently discovered superconductor, MgB$_2$. $T_c$ decreases quasi-linearly with applied pressure to 1.4 GPa at a rate of –2.0(1) K/GPa, which is somewhat larger than that derived from recently-reported *ac* susceptibility measurements. The reduction of $T_c$ is consistent with the BCS picture, in a similar way to the C$_{60}$-based superconductors. Taking into account the pressure dependence of the unit cell volume, $V$, the volume coefficient of $T_c$, $d(\ln T_c)/dV$ is significantly large.

The discovery of superconductivity in magnesium diboride with the remarkably high transition temperature of 39 K has generated considerable interest [1]. MgB$_2$ adopts a very simple primitive hexagonal crystal structure, comprising interleaved two-dimensional boron and magnesium layers [2]. Assuming full charge-transfer from Mg to the boron 2D sheets, the latter are both isostructural and isoelectronic to the graphene sheets. The appearance of high-$T_c$ superconductivity in such a simple system, which does not contain transition metals has led to considerable optimism for raising $T_c$ to higher values and has generated exciting arguments on the possible mechanism of superconductivity. In particular, a phonon-mediated BCS-type mechanism is a strong candidate [3] with superconductivity principally associated with the metallic boron 2D sheets and the high transition temperature being ensured by strong electron-phonon interactions and large vibrational energies for the light B atoms. Alternatively, as MgB$_2$ is hole-doped [4], superconductivity may be understood within the hole undressing formalism developed for the high-$T_c$ cuprate superconductors [5]. Support for the

BCS-type mechanism has derived from measurements of the boron isotope effect ($\Delta T_c$= 1.0 K, partial boron isotope exponent $\alpha_B$~ 0.26) [6] and the temperature dependence of the superconducting energy gap, $\Delta(T)$ [7]. The value of the ratio, $(2\Delta(0)/kT_c)$ has been reported both as ~3 from tunneling [7] and as ~5 from $^{11}$B NMR [8] measurements, implying weakly and strongly coupled superconductivity, respectively. A key difference between the currently competing models of superconductivity relates to the effect of applied pressure on $T_c$ with the phonon-mediated and the hole undressing mechanisms predicting negative and positive pressure coefficients, respectively. *ac* susceptibility measurements have found decreasing values of $T_c$ with applied pressure at the rate of $-dT_c/dP$~ 1.6 K/GPa up to 1.84 GPa [9]. Here we report an investigation of the conducting properties of MgB$_2$ as a function of pressure to 1.4 GPa. We similarly find a negative pressure coefficient but of somewhat larger magnitude, 2.0(2) K/GPa.

A pelletised mixture of stoichiometric amounts of Mg chips and boron powder was loaded in a stainless steel tube, which was then sealed in a quartz tube and heated at 800°C for several hours. The quality of the as-grown pellet sample was checked by powder X-ray diffraction using synchrotron radiation at Spring-8, Japan. A *dc* magnetization measurement with a commercial SQUID magnetometer at 10 Oe showed a sharp transition at 39 K with more than 100% shielding at 5 K before any correction for demagnetization effects. A standard four-probe resistivity measurement was made on an as-grown pellet sample. Hydrostatic pressure was generated using a Swenson-type Be-Cu piston-cylinder apparatus. Fluorinert FC70 was used as a pressure-transmitting medium. The sample temperature was recorded with a platinum-cobalt thermometer, embedded in the high-pressure cell. Since the piston-cylinder apparatus allows the application of pressure only at room temperature, the present system was calibrated for the pressure drop, which occurs on cooling using K$_3$C$_{60}$ and Pb samples. The system has proven quite robust in that an applied pressure of 0.35 GPa at room temperature was repeatedly found to be equivalent to ambient pressure at 20 K. The estimated error of the pressure was on the order of ±0.05 GPa.

Fig. 1 shows the temperature dependence of the resistivity at various pressures. The resistivity at room temperature was 0.15 mΩ·cm, showing metallic behaviour but with a rather small residual resistance ratio (RRR) of about 2. Superconductivity was observed below 39 K with a somewhat broad onset. Two values of the transition temperature were defined at the mid-point of the resistivity drop and at zero resistance. $T_c$ is found to decrease smoothly upon application of pressure. The relation between $T_c$ and pressure is summarised in Fig. 2, which clearly shows that $T_c$ shifts essentially linearly with pressure in this range. The rate $dT_c/dP$ was determined as −1.9(1) K/GPa and −2.14(6) K/GPa for mid-point and zero-resistance, respectively, giving an average value of 2.0(1) K/GPa. These are slightly larger than the value derived by $ac$ susceptibility measurements (−1.6 K/GPa) [9].

Of paramount importance is the determination of the underlying mechanism for superconductivity in $MgB_2$ and experimental data are rapidly being accumulated. A BCS-type pairing interaction mediated by high-frequency boron phonon modes is consistent with the reduction of $T_c$ under pressure and indicates that, within the BCS picture, the reduction of the density of states at the Fermi energy, $N(\varepsilon_F)$ due to the contraction of the B-B and Mg-B interatomic distances dominates the hardening of phonon frequencies that could cause increase of $T_c$, as external pressure is applied. This situation is reminiscent of other high-$T_c$ BCS-type superconductors, like the alkali metal intercalated fullerides, in which high frequency carbon-based intramolecular phonons mediate electron pairing [10].

At first glance, the normalized pressure coefficient of $d\ln T_c/dP = -0.05$ GPa$^{-1}$ in $MgB_2$ is an order of magnitude smaller than that in the $K_3C_{60}$ superconductor, −0.5 GPa$^{-1}$ [11]. However, the relevant comparison ought to be between the volume or lattice parameter pressure coefficients of $T_c$. $K_3C_{60}$ is essentially a fairly soft molecular material with relatively weak intermolecular interactions leading to a high volume compressibility, $d\ln V/dP = -0.036$ GPa$^{-1}$ [12], while $MgB_2$ is a stiff tightly-packed incompressible solid with $d\ln V/dP = -0.0083(3)$ GPa$^{-1}$ [13]. Taking these data into account, we obtain $d(\ln T_c)/dV = 0.22$ Å$^{-3}$, a value three times larger than that of fulleride superconductors (~0.07 Å$^{-3}$), implying an even greater sensitivity of the superconducting properties of $MgB_2$ to the interatomic distances. This sensitivity is also evident if the lattice parameter coefficients are considered as for both intraplane and interplane directions ($d(\ln T_c)/da = 7.2$ Å$^{-1}$ and $d(\ln T_c)/dc = 4.5$ Å$^{-1}$, respectively), these are significantly larger than the value of 1.7 Å$^{-1}$ for $K_3C_{60}$.

In conclusion, the observed negative pressure coefficient of $T_c$ in the $MgB_2$ superconductor assumes a very large value, consistent, for a rigid-band picture, with an extremely sensitive dependence of $N(\varepsilon_F)$ on the B-B and Mg-B bonding distances within the framework of BCS theory. We note however that for non-rigid band behaviour and depending on the extent of doping in $MgB_2$, other competing mechanisms may still be able to account for the observed experimental data.

Figure Captions

Figure 1
Resistance normalised to the 40 K values *vs* temperature for $MgB_2$ at several applied pressures.

Figure 2
Pressure dependence of $T_c$ determined by the mid-point (filled circles) and zero-resistance (open circles).

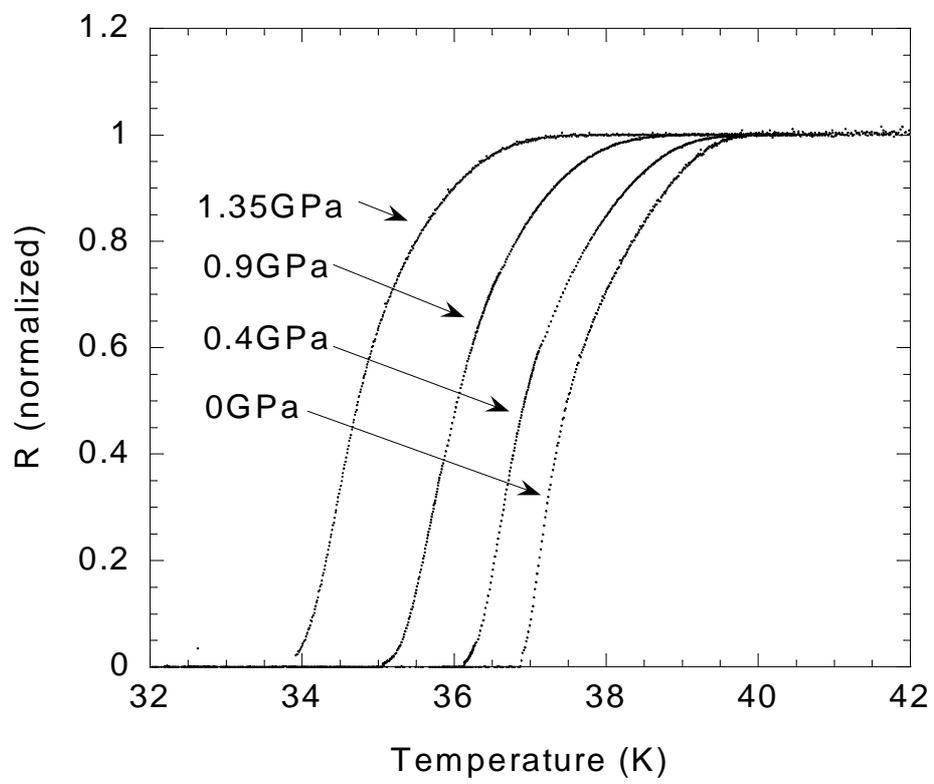

Figure 1    E Saito et al.

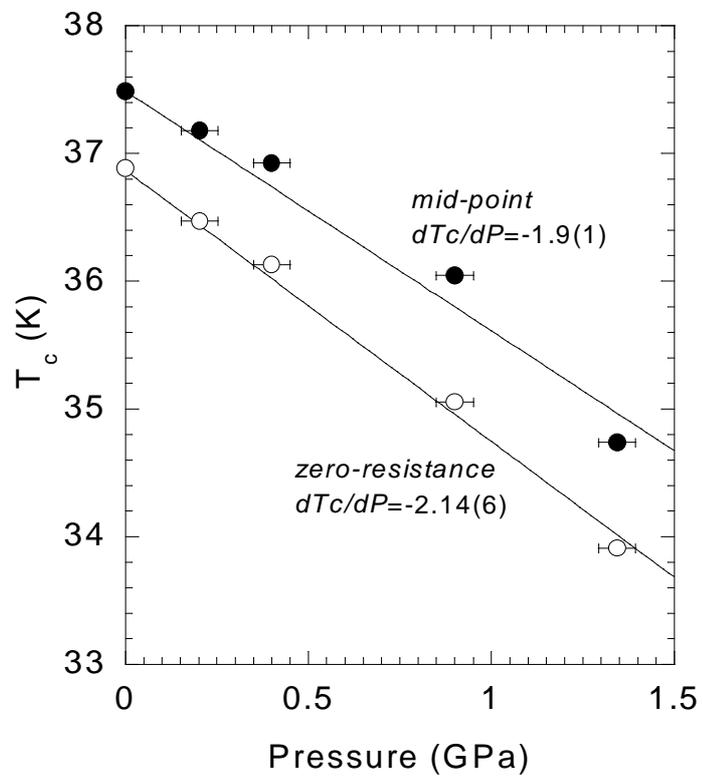

Figure 2    E Saito et al.